\documentstyle[aps,prl,multicol,psfig]{revtex}
\begin{document}
\draft
\title{Strong exciton binding in quantum structures \\
through remote dielectric confinement}

\author{Guido Goldoni, Fausto Rossi, and Elisa Molinari}

\address{Istituto Nazionale per la Fisica della Materia (INFM), and \\
Dipartimento di Fisica, Universit\`a di Modena, Via Campi
213/A, I-41100 Modena, Italy}

\date{\today}
\maketitle

\begin{abstract}

We propose a new type of hybrid systems formed by conventional
semiconductor nanostructures with the addition of remote insulating
layers, where the electron-hole interaction is enhanced by combining
{\em quantum} and {\em dielectric confinement} over different length
scales. Due to the polarization charges induced by the dielectric
mismatch at the semiconductor/insulator interfaces, we show that the
exciton binding energy can be more than doubled. For conventional
III-V quantum wires such remote dielectric confinement allows exciton
binding at room temperature.

\end{abstract}


\begin{multicols}{2}

\narrowtext

The electron-hole Coulomb interaction in semiconductors leads to bound
excitonic states that could play a crucial role in the next generation
of optical devices. For this purpose, however, the binding energy
$E_b$ must exceed the thermal energy at room temperature, a condition
that is not yet met in conventional III-V materials. In fact, a large
enhancement of $E_b$ with respect to bulk materials has been obtained
by confining electron and hole wavefunctions in nanostructures of low
dimensionality (quantum confinement), the most promising type of
structures being quasi-one-dimensional systems (quantum wires); within
GaAs-based materials, however, the observed values of $E_b$ are still
well below the room-temperature thermal energy.

In this letter we propose an alternative approach to enhance $E_b$,
that combines {\em quantum confinement} with {\em remote dielectric
confinement}. As first pointed out by Keldysh \cite{Keldysh}, the
electron-hole Coulomb attraction can be greatly enhanced in layered
structures with strong dielectric mismatch, due to the polarization
charge induced at the interfaces. For conventional semiconductor
nanostructures such as GaAs/AlGaAs- or GaAs/InGaAs-based samples,
however, this is a minor effect due to the small dielectric mismatch
between the constituents \cite{Andreani}. On the other hand,
interfaces between III-V semiconductors and materials with very
different dielectric constants, such as oxides, are usually very far
from the excellent optical quality of the conventional ones. Our
approach is based on the idea that {\em quantum and dielectric
confinement can be spatially separated}, since they are effective over
different length scales. We will show that a very large increase in
$E_b$ can be obtained in GaAs-based quantum wires by adding remote
insulating layers: these induce strong dielectric confinement without
degrading the good optical properties ensuing from quantum
confinement.

We consider GaAs/AlGaAs quantum wires (QWI), where
excitonic properties 
have been studied extensively \cite{reviews}. We focus on QWIs
with T-shaped \cite{T-growth} 
and V-shaped \cite{V-growth} cross sections, which rank among 
the best available samples from the point of view of optical properties. 
Starting from these geometries, we design hybrid structures
adding oxide layers at some distance from the 
QWI; In the new structures, the electron and hole 
wavefunctions are confined by the inner, 
lattice-and-dielectric-matched GaAs/AlGaAs
interfaces; the outer AlGaAs/oxide interfaces,
owing to the strong dielectric mismatch, provide polarization 
charges thus enhancing the electron-hole interaction. 

To obtain quantitative predictions of excitonic properties 
in such hybrid structures, an accurate description 
of the complex interplay between quantum and dielectric confinement is 
needed.
To this end, we have developed a novel theoretical scheme that 
allows to treat arbitrary dielectric configurations, 
where the low symmetry makes simple image-charge methods 
\cite{Andreani,Kumagai} not applicable. 
Furthermore, we adopt a non-perturbative approach for the
self-energy term, which, in principle, is needed to describe 
strong dielectric mismatch combined with shallow 
confining potentials, as in some state-of-the-art QWIs.
The electron-hole interaction is treated within the conventional 
approach of the semiconductor Bloch equations, adapted to quasi 
one-dimensional (1D) systems \cite{Rossi}.

More specifically, for a spatially modulated dielectric constant
$\epsilon({\bf r})$ the Coulomb
interaction
between two charges of opposite sign
$\pm e$ sitting at positions
${\bf r}$ and ${\bf r}'$ ---our electron-hole pair---
is given by $V({\bf r},{\bf r}')=-e^2 G({\bf r},{\bf r}')$, where
$G({\bf r},{\bf r}')$  
is  the Green's function of the Poisson operator, i.e.,
\begin{equation}
\bbox{\nabla}_{\bf r}\cdot\epsilon({\bf r}) \bbox{\nabla}_{\bf r}
G({\bf r},{\bf r}')= -\delta({\bf r}-{\bf r}').
\label{Poisson}
\end{equation}
We see that the space dependence of $\epsilon({\bf r})$
modifies $G({\bf r},{\bf r}')$ with respect to the homogeneous
case, where $\epsilon({\bf r})=\epsilon_\circ$ implies
$G_\circ({\bf r},{\bf r}') = {1\over 4\pi\epsilon_\circ 
|{\bf r}-{\bf r}'|}$.
This, in turn, gives rise to significant modifications in the excitonic 
spectrum; within the conventional Hartree-Fock scheme, such 
modifications originate from the electron-hole Coulomb matrix 
elements entering the evaluation of the absorption spectrum
\cite{Rossi}:
\begin{equation}
V^{eh}_{ij} = -e^2 \!\! \int \! \Phi_{i}^{e*}({\bf r}) 
\Phi_{j}^{h*}({\bf r}')
G({\bf r},{\bf r}') \Phi_{i}^h({\bf r}') \Phi_{j}^e({\bf r})
d{\bf r} d{\bf r}'.
\label{CME}
\end{equation}
Here, $\Phi^{e(h)}_l$ denotes the  electron (hole) single-particle
envelope function for the $l \equiv k_z,\nu$ state of the QWI, 
$k_z$ the wavevector along the wire, and $\nu$ the subband index.

To study realistic geometries, we find it convenient to 
cast the problem in Fourier space, following the theoretical scheme
in Ref.\ \onlinecite{Rossi}.
It is easy to rewrite Eqs. (\ref{Poisson}) and (\ref{CME})
(the symbol $\widetilde{\ }$ denotes Fourier 
transform throughout):
\begin{equation}
\sum_{{\bf k}^{''}}\widetilde{\epsilon}({\bf k}-{\bf k}^{''}) 
{\bf k}\cdot{\bf k}^{''}
\widetilde{G}({\bf k}^{''},{\bf k}') = \delta({\bf k}+{\bf k}'),
\label{FTPoisson}
\end{equation}
\begin{equation}
V^{eh}_{ij} = -e^2
\sum_{{\bf k}} F^e_{ij}({\bf k}) \sum_{{\bf k}'} 
\widetilde{G}({\bf k},{\bf k}')  
F^h_{ji}({\bf k}'),
\label{FTCME}
\end{equation}
where $F^{e/h}_{lm}({\bf k}) = \int 
\Phi^{e/h*}_{l}({\bf r})e^{i{\bf k}\cdot{\bf r}}\Phi^{e/h}_{m}({\bf
r})d{\bf r}$.  
We stress that, in order to evaluate the Coulomb matrix elements 
in (\ref{FTCME}), it is not necessary to solve 
Eq.\ (\ref{FTPoisson}) for each $\widetilde{G}({\bf k},{\bf k}')$. 
In fact, if we multiply (\ref{FTPoisson}) by
$F^h_{ji}({\bf k}')$ and sum over ${\bf k}'$, 
we get a Poisson equation for the ``potential'' 
$v_{ji}({\bf k}) = \sum_{{\bf k}'} \widetilde{G}({\bf k},{\bf k}') 
F^h_{ji}({\bf k}')$ [see Eq.\ (\ref{FTCME})]
arising from the ``source'' $F^h_{ji}({\bf k}')$, which is
solved only once for each pair $i,j$. 
 
The Green's function $G({\bf r},{\bf r}')$ also gives rise to a 
self-energy term 
\begin{equation}
\Sigma({\bf r}) = \frac{e^2}{2} \lim_{{\bf r}'\rightarrow{\bf r}}
\left[G({\bf r},{\bf r}')-G_{B}({\bf r},{\bf r}')\right],
\label{SE}
\end{equation}
where $G_{B}({\bf r},{\bf r}')={1\over 4\pi\epsilon({\bf r})|{\bf
r}-{\bf r}'|}$
is the (local) bulk solution of (\ref{Poisson}).
$\Sigma$ is a local correction (equal for
electrons and holes) which adds to the confining potential in 
determining the single-particle envelope functions $\Phi^{e/h}_l$. 
This self-energy contribution is 
evaluated within the same plane-wave approach by solving 
(\ref{FTPoisson}) at a set of ${\bf k}'$~\cite{nota-k}:
it can be shown that its Fourier transform is
\begin{equation}
\widetilde{\Sigma}({\bf G}) = \frac{e^2}{2}
\sum_{\bf g} \left[\widetilde{G}({\bf G}+\frac{{\bf g}}{2},{\bf
G}-\frac{{\bf g}}{2})-
           \widetilde{G}_{B}({\bf G},{\bf g})\right],
\label{FTSE}
\end{equation}
where $\widetilde{G}_B({\bf G},{\bf g})=\widetilde{\epsilon^{-1}}({\bf
G})/g^2$ 
with the definitions ${\bf G}=({\bf k}+{\bf k}')/2$ and ${\bf g}={\bf
k}-{\bf k}'$ 
\cite{notaSE2}.

We first discuss our findings for QWIs 
obtained by epitaxial growth on V-grooved substrates 
(V-QWIs) \cite{V-growth}. 
As a reference sample, we consider a GaAs wire
with AlAs barriers \cite{Rinaldi94}, and we add two oxide layers, 
below and above the QWI [see Fig.\ 1(a)], at a 
distance $L$ from the GaAs/AlAs interfaces \cite{nota-ossido}. 
Note that the oxide layers are characterized by a small dielectric 
constant, that we take equal to 2 \cite{Fiore}. 

Our results for the V-QWRs are shown in Fig.\ 1. For the sample shown
in Fig.\ 1(a), we find $E_b=29.3\,\mbox{meV}$, to be compared with
$13\,\mbox{meV}$ of the conventional (i.e., with no oxide layers)
structure. Fig.\ 1(a) shows that the origin of this dramatic
enhancement is the large polarization of the AlAs/oxide interfaces
induced by the hole charge density\cite{nota-eh-symm}; the polarization 
is larger in the
region where the hole is localized, and is more pronounced at the
lower interface, due to the larger curvature. A small polarization
charge is also induced at the GaAs/AlAs interface, due to the small
dielectric mismatch. Note that quantum confinement localizes the
wavefunction well within the inner interfaces; therefore, the
AlAs/oxide interface does not affect the electron and hole
wavefunctions.

In Fig.\ 1(b) we show the calculated $E_b$ for selected values of $L$.
We also show, for comparison, the calculated binding energy
for the conventional
structure, $E_0$, and the room-temperature thermal energy. $E_b$ is
maximum
when the oxide layer is at minimum distance \cite{limit}, $L=0$:
it is enhanced by a factor larger than 3 with respect to $E_0$, 
and it is well above $kT_{\mbox{\tiny room}}$ \cite{exchange}.
It is important to note that $E_b$ decreases slowly with $L$, 
and is still significantly larger than $kT_{\mbox{\tiny room}}$ 
at $L=6\,\mbox{nm}$. 
Since $E_b$ is the result of the 
Coulomb interaction of, say, the electron with the hole {\em and } the 
polarization charge which is excited at a distance $\sim L$, 
we intuitively expect $E_b$ to decay as $L^{-1}$, with a
typical decay length $L_0$ comparable to the Bohr radius 
in the QWI \cite{Rossi}; this, in turn, is of the order of the
confinement length. Indeed, Fig.\ 1(b) shows that $E_b$ is very well
interpolated by 
\begin{equation}
E_b(L) = E_0 + \frac{E_b(0)}{1+L/L_0}
\label{fit}
\end{equation}
with $L_0=6.56\,\mbox{nm}$. Note that $E_b$ crosses $kT_{\mbox{\tiny
room}}$ when $L$ 
is as large as 9 nm. 

A second type of structures, which have recently attracted
considerable attention, are the so-called T-shaped wires (T-QWI),
obtained by the cleaved-edge overgrowth method
\cite{reviews,T-growth}. 
The typical
sample of our calculations [see Fig.\ 2(a)] consists of a T-QWI with
GaAs parent quantum wells (QWs) of the same width, and AlAs barriers.
An oxide layer is added on top of the exposed surface at a distance
$L$ from the underlying QW. Note that, in this case, an oxide layer is
present only on one side of the QWI. As in the case of the V-QWIs, a
strong polarization charge forms at the AlAs/oxide interface, with a
maximum in the region of the hole wavefunction confinement. A small
polarization charge is also present at the GaAs/AlAs interface, peaked
around the corners of the intersecting QWs.
In Fig.\ 2(b) we show the calculated $E_b$ vs $L$. The
binding energy for the conventional structure, $E_0$,
and the room-temperature thermal energy are also shown for comparison.
As in the case of the V-QWI, $E_b$
is maximum at $L=0$, where it is enhanced by a factor of 1.5 with
respect to $E_0$, and decreases slowly with $L$.
Although $E_b$ is smaller than in the case of the V-QWI studied
previously,
for the smallest $L$ values $E_b$ is still of the order of
$kT_{\mbox{\tiny room}}$. 
It is important to note that the reduced effect of dielectric
confinement
in the T-QWI sample with respect to the previous example of V-QWI 
is just due to the presence of a single oxide layer,
i.e., geometric effects due to different cross-sections 
play a minor role. In fact,
despite the very different geometry, $E_b$ decays with $L$ in the same
way 
in both cases. As shown in Fig.\ 2(b), also in the case of T-QWIs 
$E_b$ is very well interpolated by 
Eq.\ (\ref{fit}) with $L_0=7.55\,\mbox{meV}$, which is still of the
order 
of the QWI confinement length. 
Note that the above examples of structures are based on standard 
state-of-the-art QWIs that were previously studied in the 
literature~\cite{T-growth,V-growth}, and no particular optimization 
of $E_b$ with respect to sample 
parameters (constituents and/or geometry) has been attempted. 

Finally, we discuss the effect of the self-energy term on $E_b$. We
have compared the full calculations discussed above with calculations
performed neglecting $\Sigma({\bf r})$ in the single-particle
potential. For the V-QWI we have verified that the self-energy
contribution tends to increase $E_b$, but it is so small
($<0.2\,\mbox{meV}$) that the two results cannot be distinguished on
the scale of Fig.\ 1(b). In the case of the T-QWIs, on the other hand,
the self-energy contribution is qualitatively and quantitatively
different, as can be seen from Fig.\ 2(b); in this case, in fact, 
it amounts to $\sim 1\,\mbox{meV}$ at the
smallest $L$, and tends to reduce $E_b$. This is a consequence of the
interplay between the dielectric confinement and the shallow quantum
confinement of these structures; this is apparent from Fig.\ 3, where
we compare the electron and hole single-particle wavefunctions
calculated neglecting [Fig.\ 3(a)] and including [Fig.\ 3(b)] the
self-energy term. The self-energy potential [Fig.\ 3(c)] inside the
GaAs layer pushes the electron and hole wavefunctions away from the
oxide layer; due to the different masses and shallow confining
potentials, such shift is different for electrons and holes, and the
overlap is diminished, thereby reducing $E_b$.

In summary, we have developed a theoretical scheme 
that allows to include dielectric confinement and self-energy effects 
in a full three-dimensional description of correlated electron-hole
pairs
\cite{nanocrystals}.
Our calculations show that a dramatic enhancement of the
exciton binding in GaAs-based quantum structures is made possible by
remote insulating layers, bringing $E_b$ in the range of the
room-temperature thermal energy.  This enhancement 
scales slowly with the distance of the insulating layer from the 
quantum confinement region, thus allowing to design nanostructures
that should be compatible with excellent optical efficiency.
Dielectric confinement by remote insulating layers is predicted to
be a novel powerful tool for tailoring excitonic confinement in 
semiconductor nanostructures.

We are grateful to E. Kapon, L. Sorba, and W. Wegscheider for useful
discussions.

\end{multicols}

\widetext

\newpage

\begin{figure}

\noindent
\unitlength1mm
\begin{picture}(180,110) 
\put(30,0){\psfig{figure=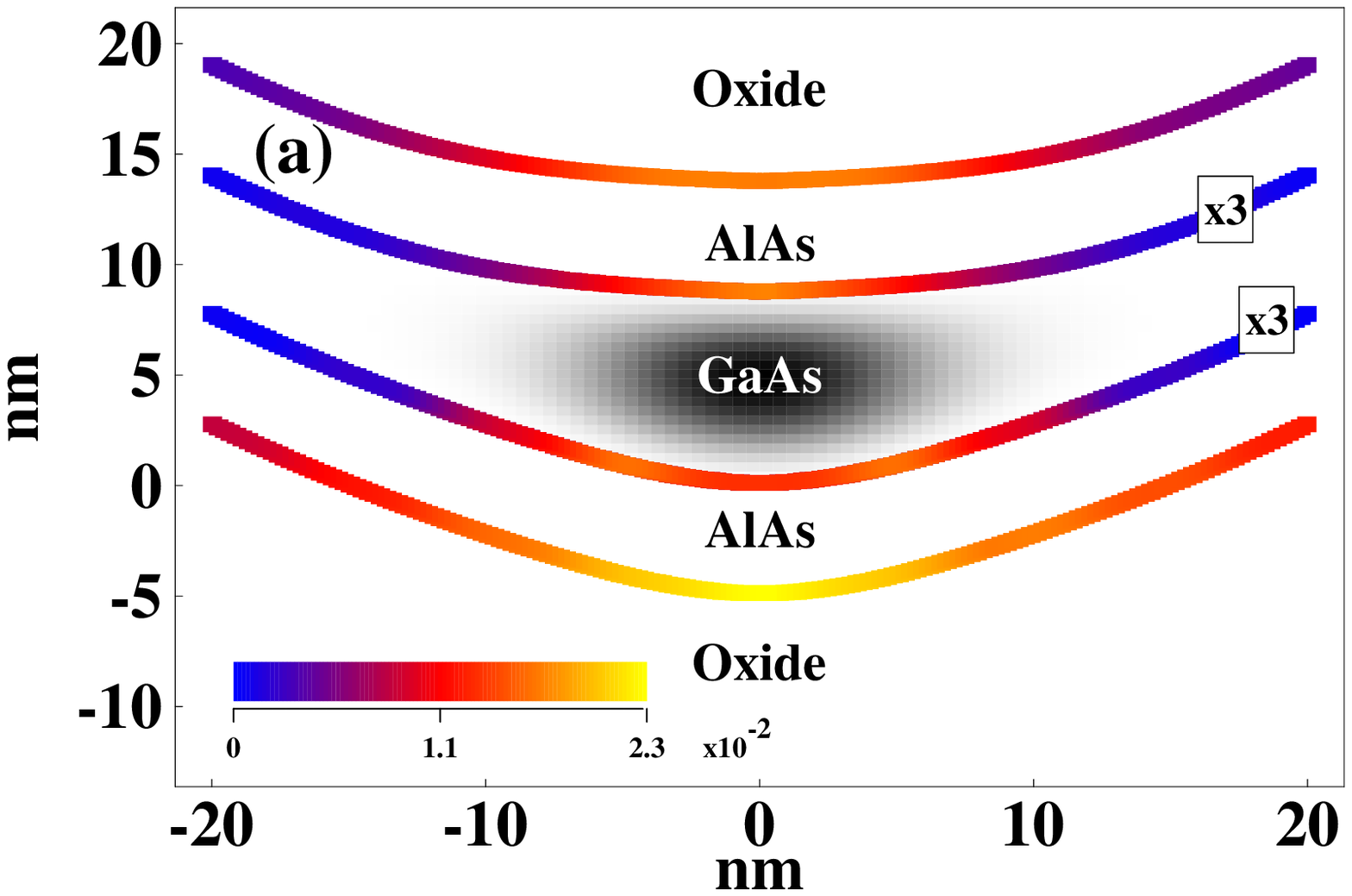,width=100mm}}
\end{picture}

\noindent
\unitlength1mm
\begin{picture}(180,110) 
\put(33,0){\psfig{figure=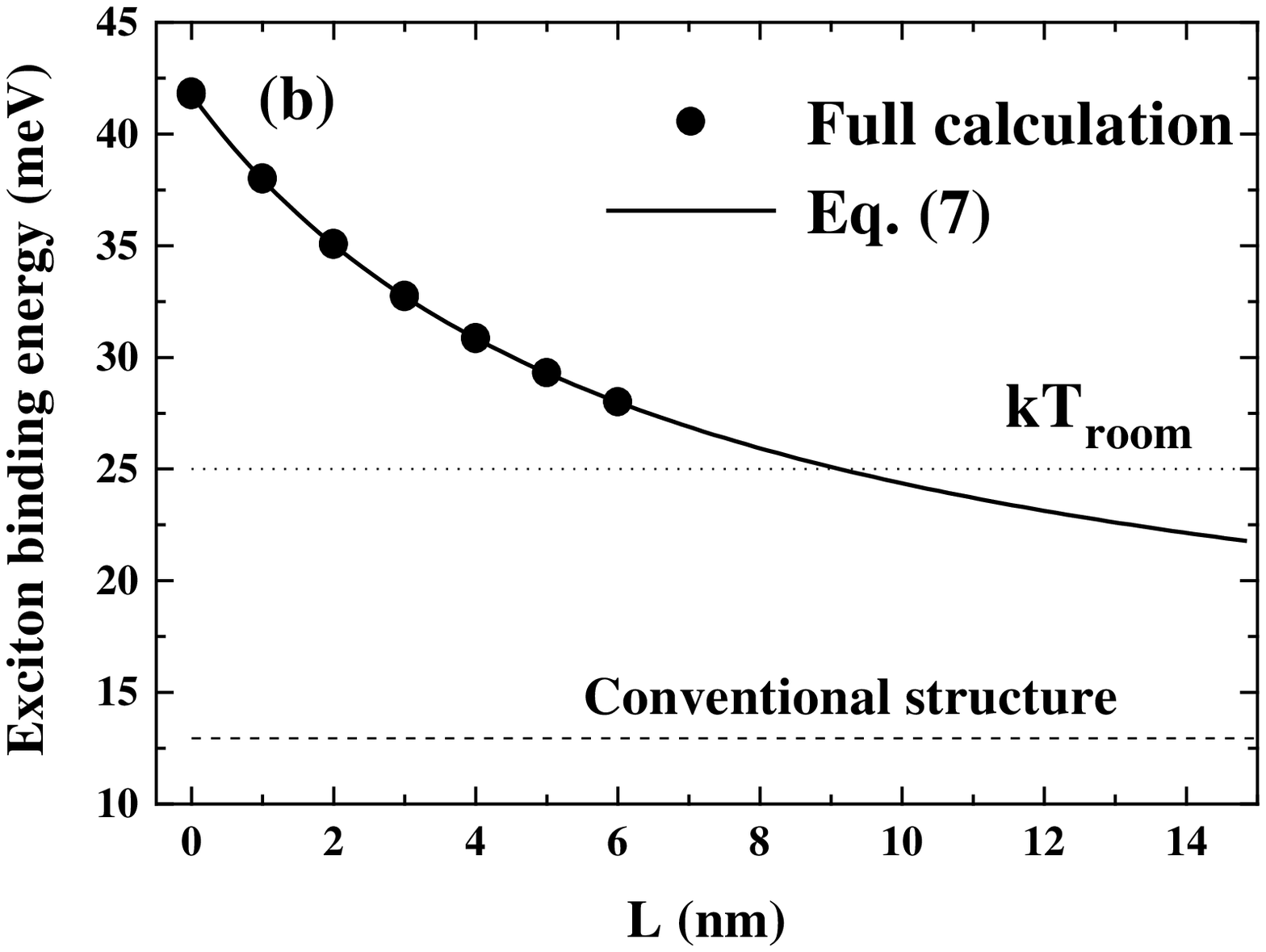,width=100mm}}
\end{picture}

\vspace{-5truecm}

\caption{(a) Cross section of a hybrid V-QWI, showing the
interface polarization charge (colors, units of nm$^{-1}$)
induced by the charge-density
distribution of the lowest-subband hole (grey-scale, arbitrary units).
The profile of the GaAs/AlAs interfaces is obtained from
Ref.\ \protect\cite{Rinaldi94};
the oxide layers are at $L=5\,\mbox{nm}$ from the GaAs/AlAs interfaces.
The polarization charge at the GaAs/AlAs interfaces
is multiplied by 3.
(b) $E_b$ versus distance of the oxide layers
from the internal interfaces, $L$.
Solid dots: full calculation. Solid line:
Eq.\ (\ref{fit}) with $L_0=6.56$ nm. Dashed line:
energy $E_0$ of the corresponding conventional structure
(no oxide layers). 
Dotted line: thermal energy at $T_{\mbox{\tiny room}}= 300\,\mbox{K}$.}
\end{figure}

\newpage

\begin{figure}

\noindent
\unitlength1mm
\begin{picture}(180,110) 
\put(30,0){\psfig{figure=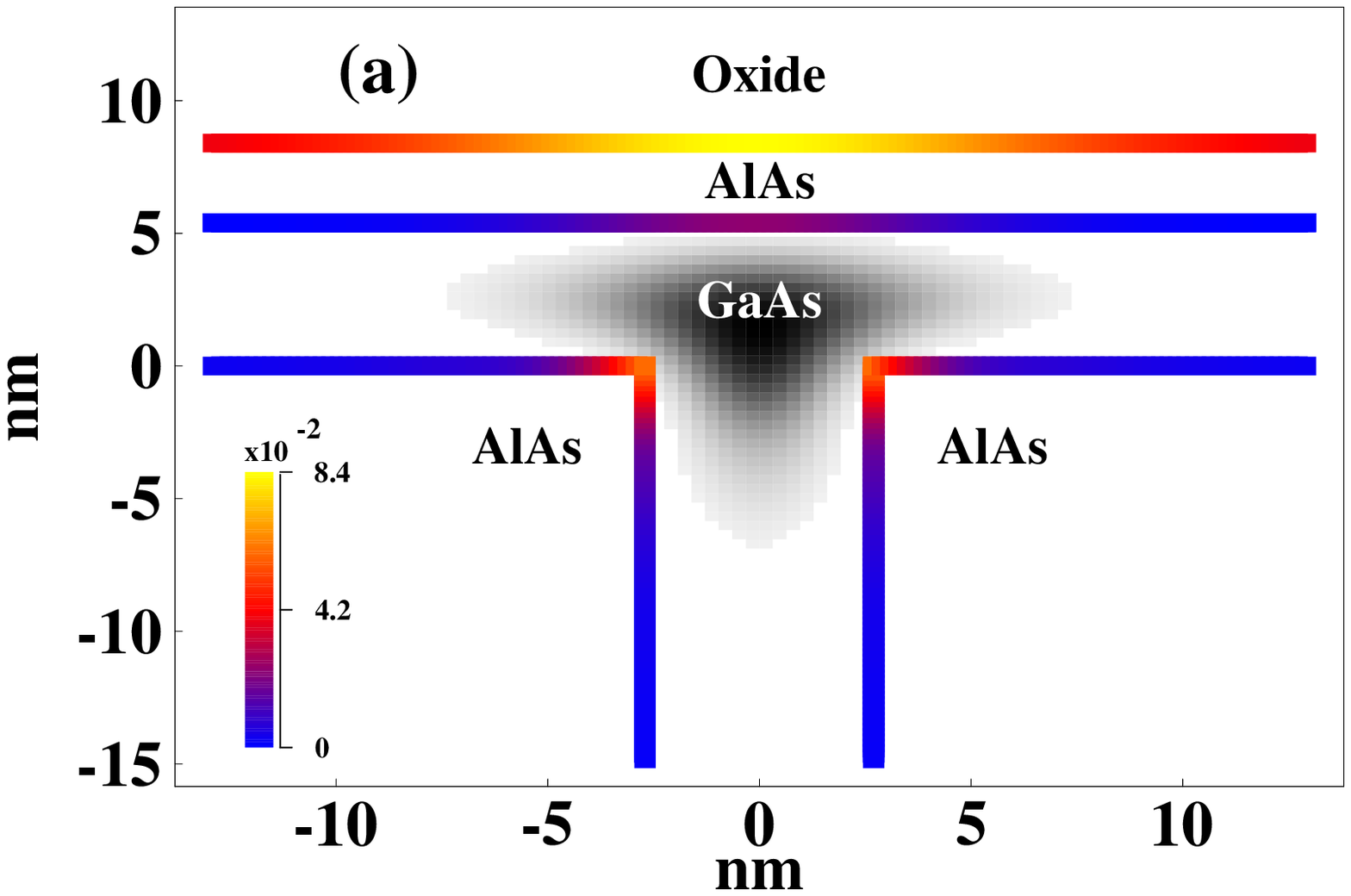,width=100mm}}
\end{picture}

\noindent
\unitlength1mm
\begin{picture}(180,110) 
\put(35,0){\psfig{figure=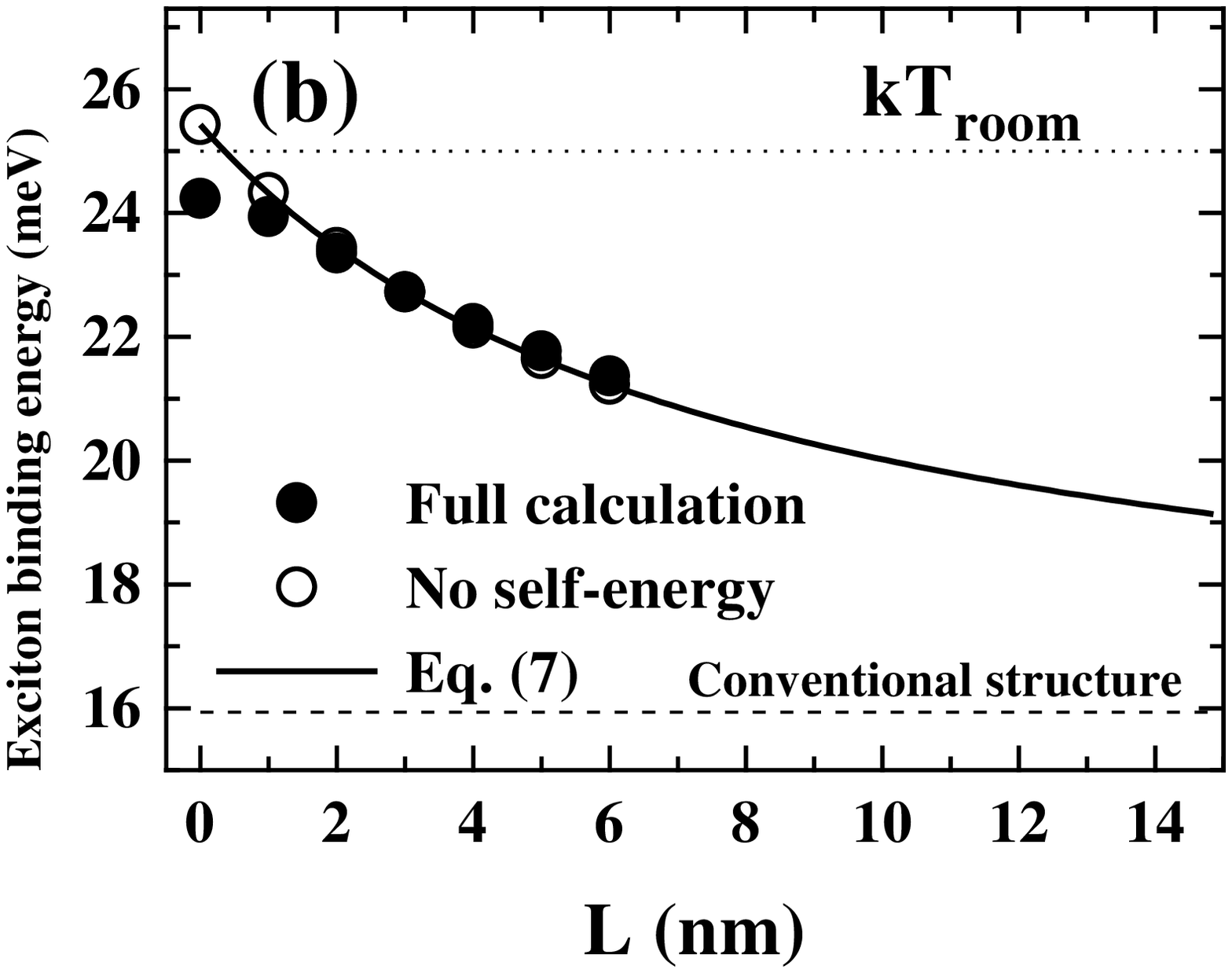,width=100mm}}
\end{picture}

\vspace{-5truecm}

\caption{
(a) Same as Fig.\ 2(a) for a hybrid T-QWI. The QW widths are
$5.4\,\mbox{nm}$; the oxide layers are at $L=3\,\mbox{nm}$ 
from the GaAs/AlAs interface. 
(b) $E_b$ versus $L$ including (solid dots) and
neglecting (empty dots) the self-energy contribution. Solid line: 
Eq.\ (\ref{fit}) with $L_0=7.55$ nm. Dashed line: 
energy $E_0$ of the corresponding conventional structure 
(no oxide layers). Dotted line: thermal energy at 
$T_{\mbox{\tiny room}}= 300\,\mbox{K}$.}
\end{figure}

\newpage

\begin{figure}

\noindent
\unitlength1mm
\begin{picture}(180,180) 
\put(15,0){\psfig{figure=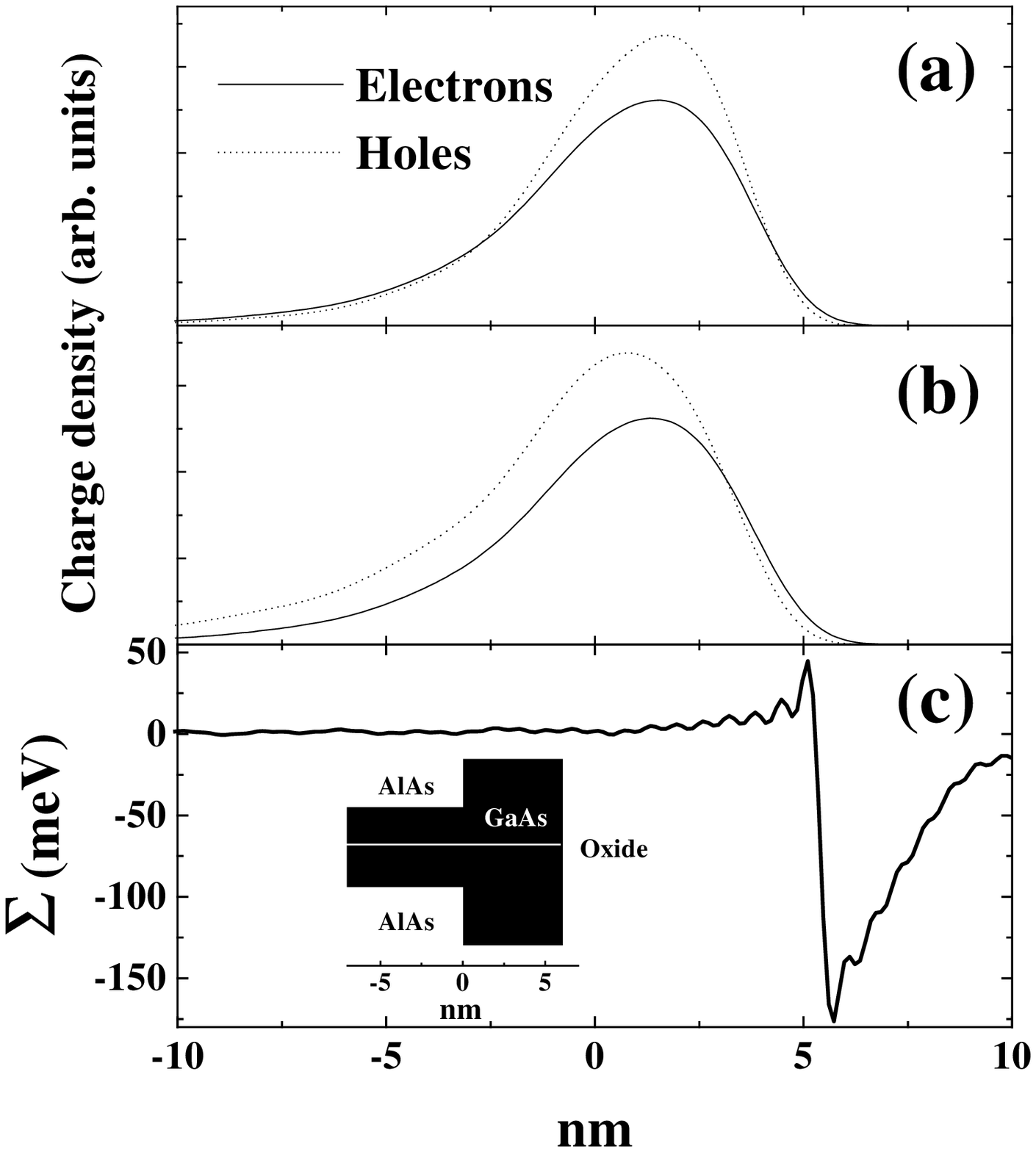,height=180mm}}
\end{picture}

\vspace{-2truecm}

\caption{Lowest-subband electron and hole charge densities for a T-QWI
with $L=0$ [see inset], neglecting (a) and including (b)
the self-energy potential $\Sigma ({\bf r})$ shown in (c). 
All curves are calculated 
along the white line shown in the inset.}
\end{figure}


\begin{references}
\bibitem{nota-ossido} 
The GaAs/AlAs band offsets are 1.036 eV (electrons) and 0.558 eV
(holes). 
Effective masses are 0.067 (electrons) and 0.38 (holes). Dielectric
constants 
are 12.5 (GaAs), 10.0 (AlAs), and 2 (oxide).
We neglect the variation of other band parameters between 
AlAs and oxide layers. Due to the exponential decay of the wavefunctions
in 
the AlAs layers, these effects should be negligible when $L$ 
exceeds a few nm.
\bibitem{nota-eh-symm} Of course there is no electron-hole asymmetry, 
and we could equivalently show the electron charge density 
and its induced polarization charge; the present
choice is consistent with the choice of $F^h_{ji}$ as ``source'' term,
as 
discussed below Eq.\ (\protect\ref{FTCME}).
\bibitem{limit} For completeness, our calculations have been extended
down to $L=0$, although in this limit they should be taken
with caution, due to the neglect of the AlAs/oxide interface effects. 
See also note \protect\cite{nota-ossido}.
\bibitem{exchange} In spite of such a dramatic enhancement, $E_b$ is
still much smaller than the GaAs band gap, so that the usual
approximation which neglects the interband exchange interaction is
still valid in the present structures.
\end{references}
\end{document}